%% file: main.tex
\documentclass[aps,prl,twocolumn,superscriptaddress,groupedaddress,preprintnumbers]{revtex4-2} 
\usepackage{graphicx}  
\usepackage{subcaption}
\usepackage{multirow}
\usepackage{dcolumn}   
\usepackage{bm}        
\usepackage{amssymb}   
\usepackage{amsmath}
\usepackage{mathtools}
\usepackage{cases}
\usepackage{mathrsfs}
\usepackage{color}
\usepackage{hyperref}  
\usepackage{orcidlink}
\usepackage{CJK}
\usepackage{caption}
\usepackage{ulem}
\usepackage{array}
\usepackage{xcolor}
\usepackage{tikz}

\definecolor{color97-1}{RGB}{94,129,181}
\definecolor{color97-3}{RGB}{143,176,50}

\newcommand{\be}{\begin{equation}}
\newcommand{\ee}{\end{equation}}
\newcommand{\bea}{\begin{eqnarray}}
\newcommand{\eea}{\end{eqnarray}}

\definecolor{mycyan}{RGB}{0,255,255}
\definecolor{myorange}{RGB}{255,192,6}
\newcommand{\solidblack}{%
\begin{tikzpicture}[baseline=-0.6ex]
\draw[black, line width=1.5pt, line join=round] (0,0) -- (1.1,0);
\end{tikzpicture}}
\newcommand{\dashedblack}{%
\begin{tikzpicture}[baseline=-0.6ex]
\draw[black, line width=1.5pt, line join=round, dash pattern=on 3.0pt off 1.0pt] (0,0) -- (1.1,0);
\end{tikzpicture}}
\newcommand{\solidsteel}{%
\begin{tikzpicture}[baseline=-0.6ex]
\draw[color97-3, line width=1.5pt, line join=round] (0,0) -- (1.1,0);
\end{tikzpicture}}
\newcommand{\dashedsteel}{%
\begin{tikzpicture}[baseline=-0.6ex]
\draw[color97-3, line width=1.5pt, line join=round, dash pattern=on 3.0pt off 1.0pt] (0,0) -- (1.1,0);
\end{tikzpicture}}
\newcommand{\dottedblack}{%
\begin{tikzpicture}[baseline=-0.6ex]
\draw[black, line width=0.6pt, dash pattern=on 0.8pt off 1.2pt, line join=round, line cap=round] (0,0) -- (1.1,0);
\end{tikzpicture}}
\newcommand{\dottedsteel}{%
\begin{tikzpicture}[baseline=-0.6ex]
\draw[color97-3, line width=0.6pt, dash pattern=on 0.8pt off 1.2pt, line join=round, line cap=round] (0,0) -- (1.1,0);
\end{tikzpicture}}
\newcommand{\dashedblackthin}{
\begin{tikzpicture}[baseline=-0.6ex]
\draw[black, line width=0.6pt, line join=round, dash pattern=on 3pt off 1pt] (0,0) -- (1.1,0);
\end{tikzpicture}}
\newcommand{\solidorange}{%
\begin{tikzpicture}[baseline=-0.6ex]
\draw[myorange, line width=1.2pt, line join=round] (0,0) -- (1.1,0);
\end{tikzpicture}}
\newcommand{\dashedorange}{%
\begin{tikzpicture}[baseline=-0.6ex]
\draw[myorange, line width=0.6pt, line join=round, dash pattern=on 3pt off 1pt] (0,0) -- (1.1,0);
\end{tikzpicture}}
\newcommand{\solidcyan}{%
\begin{tikzpicture}[baseline=-0.6ex]
\draw[mycyan, line width=1.2pt, line join=round] (0,0) -- (1.1,0);
\end{tikzpicture}}
\newcommand{\dashedcyan}{%
\begin{tikzpicture}[baseline=-0.6ex]
\draw[mycyan, line width=0.6pt, line join=round, dash pattern=on 3pt off 1pt] (0,0) -- (1.1,0);
\end{tikzpicture}}

\definecolor{sigBlack}{HTML}{000000}
\definecolor{sigPurple}{HTML}{8E44AD} 
\definecolor{sigGreen}{HTML}{76FF03}  

\begin{document}
	\title{A Unified Origin of Primordial Black Hole Dark Matter and Nanohertz Gravitational Waves}

\author{Guillem Domènech${}^{1,2}$\,
\orcidlink{0000-0003-2788-884X}}
\email{Corresponding author: guillem.domenech@itp.uni-hannover.de}
\author{Shi Pi${}^{3,4,5}$\,
\orcidlink{0000-0002-8776-8906}}
\email{Corresponding author: shi.pi@itp.ac.cn}
\author{Ao Wang${}^{3,6,7}$\,
\orcidlink{0009-0002-6559-5212}}
\email{Corresponding author: wangao@itp.ac.cn}
\affiliation{%
    $^{1}$Institute for Theoretical Physics, Leibniz University Hannover, Appelstraße 2, 30167 Hannover, Germany
}%
\affiliation{
    $^{2}$Max-Planck-Institut für Gravitationsphysik, Albert-Einstein-Institut, 30167 Hannover, Germany
}%
\affiliation{
    $^{3}$Institute of Theoretical Physics, Chinese Academy of Sciences, Beijing 100190, China}
\affiliation{
    $^{4}$ Center for High Energy Physics, Peking University, Beijing 100871, China}
\affiliation{
    $^{5}$ Kavli Institute for the Physics and Mathematics of the Universe (WPI), The University of Tokyo, Kashiwa, Chiba 277-8583, Japan}
\affiliation{
    $^{6}$ School of Physical Sciences, University of Chinese Academy of Sciences, Beijing 100049, China}
\affiliation{
    $^{7}$ Department of Physics, Kyoto University, Kyoto 606-8502, Japan}

	    
	\date{\today}
	\begin{abstract} 
    Recent high-cadence observations by Subaru-HSC have identified a population of ultrashort-timescale microlensing events, providing a compelling window for planet-mass primordial black holes (PBHs) to constitute the entirety of dark matter. In this \textit{Letter}, we demonstrate that this PBH population and the nanohertz stochastic gravitational-wave (GW) background reported by pulsar timing arrays (PTAs) can be naturally unified by a single primordial origin: a broad, nearly-flat enhancement of the curvature power spectrum with an amplitude of $\mathcal{O}(10^{-2})$. The resulting PBH mass function spans the planet-to-solar mass range, while remaining consistent with all current observational constraints. This unified PBH--induced-GW framework makes concrete multi-messenger predictions, which can be decisively scrutinized by forthcoming microlensing surveys, next-generation PTAs, space-borne interferometers, precision astrometry, and laser ranging experiments.
\end{abstract}
	\maketitle

\textit{Introduction.}---The direct detection of gravitational waves (GWs) from black hole mergers by LIGO/Virgo \cite{Abbott:2016blz,Abbott:2016nmj,Abbott:2017vtc,Abbott:2017gyy,Abbott:2017oio,TheLIGOScientific:2017qsa} has renewed interest in primordial black holes (PBHs). Such black holes formed from the gravitational collapse of large-density perturbations in the early Universe and provide a unique probe of physics at energy scales far beyond those of terrestrial experiments. Depending on their mass, PBHs may explain various observations \cite{Carr:2020xqk,Carr:2023tpt,Byrnes:2025tji,Carr:2026hot}.

Asteroid-mass PBHs ($10^{-16}$–$10^{-12}M_\odot$) are a viable dark matter candidate and may be tested by future space-based GW detectors LISA \cite{LISA:2017pwj}, Taiji \cite{Luo:2019zal}, and TianQin \cite{Luo:2025ewp}. Solar mass PBHs could contribute to the merger rates measured by the LIGO/Virgo/KAGRA (LVK) collaboration, potentially accounting for events with masses or spins difficult to reconcile with conventional astrophysical formation channels. A detection of subsolar-mass black holes \cite{LVK:2022ydq,Prunier:2023uoo,Kacanja:2026byy} would be a major discovery, since they cannot form from standard stellar evolution. Note that there are a few potential candidates of sub-solar binary mergers \cite{Clesse:2020ghq,LVK:2022ydq,Morras:2023jvb,Prunier:2023uoo}. Taken together, these considerations place PBHs at the intersection of dark matter physics, gravitational-wave astronomy, and early-Universe cosmology.

Evidence of Planet-mass PBHs may be found by microlensing of background stars. Such ultra-short-timescale microlensing events are a primary target of surveys, including the Optical Gravitational Lensing Experiment (OGLE) and the Subaru Hyper Suprime-Cam (Subaru-HSC). Analysis of the OGLE 5-year microlensing dataset revealed a subpopulation of six ultra-short timescale events \cite{Mroz:2017mvf}, which can be explained by planet–mass PBHs with a fractional abundance of a few percent of dark matter \cite{Niikura:2019kqi}. Recently, Subaru-HSC reported 12 microlensing candidates toward Andromeda with light-curve durations shorter than 5 hours, identified over a total of 39.3 hours of observing time \cite{Sugiyama:2026kpv}. Using a hierarchical Bayesian analysis that combines light-curve information and Poisson statistics, the authors derive an allowed parameter region under the hypothesis that these candidates arise from PBH microlensing, with characteristic masses $M_\mathrm{PBH}\sim10^{-7}$–$10^{-6}~M_\odot$ and a fractional abundance $f_\mathrm{PBH}\sim0.1$–1, which is consistent with OGLE allowed region \cite{Niikura:2019kqi}. 
This implies that planet-mass PBHs could constitute the entirety of dark matter.

Producing a substantial fraction of planet-mass PBHs requires an enhanced curvature power spectrum with amplitude $\mathcal{P}_\mathcal{R}\sim 3\times10^{-2}$ at comoving wavenumber $k_p\sim10^{9}$ -- $10^{10}~\mathrm{Mpc}^{-1}$. The evolution of such large primordial fluctuations generically sources a detectable stochastic gravitational-wave background (SGWB) \cite{Tomita:1967wkp,Matarrese:1992rp,Matarrese:1993zf,Matarrese:1997ay,Noh:2004bc,Carbone:2004iv,Nakamura:2004rm,Ananda:2006af,Osano:2006ew,Baumann:2007zm,Alabidi:2012ex,Alabidi:2013wtp,Inomata:2016rbd,Orlofsky:2016vbd}, of which the energy density fraction is $\Omega_\mathrm{IGW}\sim10^{-9}$ in the frequency band $10^{-6}$ -- $10^{-5}$ Hz. This lies beyond the optimal sensitivity bands of the aforementioned space-borne interferometers, but can be probed through precision astrometry with instruments such as the Nancy Grace Roman Space Telescope \cite{Wang:2022sxn,Pardo:2023cag}, and lunar and satellite laser ranging \cite{Blas:2021mqw,Foster:2025nzf,Blas:2026xws}.

Lastly, in 2023, multiple pulsar timing array (PTA) collaborations, including NANOGrav, EPTA, PPTA, CPTA, and the IPTA, reported strong evidence for a SGWB in the nanohertz frequency band, characterized by $\Omega_\text{GW}\sim3\times10^{-8}$ at $f\sim 10^{-8}~\text{Hz}$ \cite{EPTA:2023fyk,EPTA:2023sfo,EPTA:2023xxk,Zic:2023gta,Reardon:2023gzh,Reardon:2023zen,NANOGrav:2023hde,NANOGrav:2023gor,InternationalPulsarTimingArray:2023mzf,Xu:2023wog}. Among new-physics interpretations, GWs induced by enhanced curvature perturbations have emerged as a comparatively favored cosmological source \cite{NANOGrav:2023hvm,Figueroa:2023zhu,Bian:2023dnv}.


The close alignment between the amplitudes of the nanohertz GWs and the scalar-induced GWs associated with planet-mass PBHs suggested by microlensing observations motivates us to consider a broad, nearly flat curvature power spectrum, which can naturally arise in theoretical scenarios including phase transition during inflation \cite{Kusenko:2020pcg,Sugiyama:2020roc}, ultra-slow-roll inflation with smooth transitions to subsequent slow-roll stages \cite{Starobinsky:1992ts,Yokoyama:1998pt,Garcia-Bellido:2016dkw,Cheng:2016qzb,Garcia-Bellido:2017mdw,Cheng:2018yyr,Dalianis:2018frf,Tada:2019amh,Xu:2019bdp,Mishra:2019pzq,Bhaumik:2019tvl,Liu:2020oqe,Fu:2020lob,Vennin:2020kng,Ragavendra:2020sop,Gao:2021dfi,Pi:2022zxs,Cole:2022xqc,Kubota:2023ked,Cole:2023wyx,Ragavendra:2023ret,Domenech:2023dxx,Fujita:2025imc,Tomberg:2025fku,Talebian:2025jeg,Miyamoto:2025qqm,Escriva:2025ftp}, or constant-roll inflation \cite{Kinney:2005vj,Martin:2012pe,Motohashi:2014ppa,Motohashi:2017aob,Motohashi:2017vdc,Atal:2018neu,Motohashi:2019tyj,Tomberg:2023kli,Wang:2024xdl,Kristiano:2024vst,Inui:2024sce,Inui:2024fgk,Shimada:2024eec}. The GW spectrum induced by such an enhanced curvature perturbation spans from $10^{-9}$ Hz to $10^{-5}$ Hz. Moreover, the PBH mass function sourced by such a broad curvature spectrum approximately scales as $f(M)\propto M^{-1/2}$, which peaks at the lowest mass \cite{Byrnes:2018clq,Byrnes:2018txb,DeLuca:2020ioi}. 
This allows a sufficiently flat spectrum to generate planet-mass PBHs to account for all dark matter.

In this \textit{Letter}, we show that a single flat curvature power spectrum can simultaneously account for the entirety of dark matter---via a population of planet-mass PBHs suggested by the recent Subaru-HSC microlensing events---and the nanohertz stochastic gravitational-wave background observed by PTAs. We perform a full analysis of this possibility, including updated constraints based on LIGO/VIRGO/KAGRA O3 data. All observations combined completely fix the model parameters and yield precise values for the amplitude and width of the primordial curvature spectrum, which can be tested in a decade by different observational projects.


\textit{Planet-mass PBHs as all dark matter.}---PBHs are formed by the gravitational collapse of high density peaks. If PBHs constitute the entirety of dark matter, the SGWB induced by the curvature perturbation has an energy density spectrum $\Omega_\mathrm{GW}\sim10^{-9}$ \cite{Saito:2008jc,Josan:2009qn,Assadullahi:2009jc,Bugaev:2009zh,Saito:2009jt,Bugaev:2010bb,Inomata:2018epa,Kalaja:2019uju}.
However, the PBH abundance is sensitive to many factors, like the amplitude and shape of the power spectrum \cite{Germani:2018jgr,MoradinezhadDizgah:2019wjf,DeLuca:2020ioi,Musco:2020jjb}, choice of window functions \cite{Ando:2018qdb,Young:2019osy,Yoo:2020dkz}, formation threshold \cite{Carr:1975qj,Shibata:1999zs,Harada:2013epa,Nakama:2013ica,Harada:2015yda,Musco:2018rwt,Escriva:2019phb,Young:2019osy,Musco:2020jjb}, the methodology (Press--Schechter \cite{Press:1973iz,Gow:2022jfb,Young:2024jsu,Saito:2025sny} vs. peak theory \cite{Green:2004wb,Yoo:2018kvb,Germani:2018jgr,Atal:2019cdz,Atal:2019erb,Young:2020xmk,Yoo:2020dkz,Taoso:2021uvl,Riccardi:2021rlf,Kitajima:2021fpq,Young:2022phe,Pi:2024ert}), \textit{etc}. As a complete peak-theory calculation for a top-hat spectrum is not yet developed, we resort to the Press–Schechter formalism with the density contrast on comoving slices $\delta\equiv\delta\rho/\rho\approx-(4/9)(\nabla^2/(H^2a^2))\mathcal{R}$ to estimate the PBH abundance, where $\mathcal{R}$ is the comoving curvature perturbation. We take a Gaussian window function $\widetilde{W}^2(k;R_\mathrm{s})=\exp\left(-k^2R_\mathrm{s}^2\right)$,  
where the smoothing scale $R_\mathrm{s}$ is roughly the horizon size, \textit{i.e.} $R_\mathrm{s}=1/(aH)=k_\mathrm{eq}^{-1}\sqrt{M_H/M_\mathrm{eq}}$. A region with $\delta>\delta_\mathrm{c}$ can collapse into a PBH with mass
\be\label{eq:criticalmass}
M\left(\delta;M_H\right)\approx \mathcal{K}M_H\big(\delta-\delta_{\mathrm{c}}\big)^{\gamma},
\ee
where $\mathcal{K}\approx4$, and $\gamma\approx0.36$ is the critical index \cite{Choptuik:1992jv,Evans:1994pj,Koike:1995jm,Niemeyer:1997mt,Hawke:2002rf,Musco:2008hv}. The threshold density contrast $\delta_\mathrm{c}$ ranges from $0.3$ to $0.6$ in the literature~\cite{Carr:1975qj,Harada:2013epa,Musco:2018rwt,Germani:2018jgr,Musco:2020jjb}, depending mainly on the profile of $\delta$ as well as the equation-of-state parameter $w$. 
For simplicity, we choose $\delta_\mathrm{c}\approx0.48$ for a top-hat spectrum in the radiation dominated era, following \cite{Musco:2020jjb}. A comparison with other choices of $\delta_\mathrm{c}$ is shown in the \textit{End Matter}. 
During the cosmic evolution, $w$ can deviate from $1/3$, at most up to $30\%$ during the QCD transition near $200~\mathrm{MeV}$. To account for this effect, we use the analytical expression $\delta_\mathrm{c}(w)$ derived in Ref. \cite{Escriva:2020tak} (Eq.(27) with $q=0.9$ corresponding to the top-hat spectrum), together with a varying $w(T)$ given in Ref. \cite{Carr:2019kxo}.
 
For a Gaussian density contrast $\delta$, the PBH mass function at formation is
\begin{align}\label{beta(M)1}
\beta(M)=\mathcal{K}\int^\infty_{\delta_{\mathrm{c}}}\mathrm{d}\delta\frac{\big(\delta-\delta_\mathrm{c}\big)^\gamma}{\sqrt{2\pi\sigma_\delta^2(M)}}\exp\left(-\frac{\delta^2}{2\sigma_\delta^2(M)}\right),
\end{align}
where $\sigma_\delta^2$ is the smoothed variance defined by $\sigma_\delta^2(M)\equiv\int\mathrm{d}\ln k~\mathcal{P}_\delta(k)\widetilde{W}^2(k,R_\mathrm{s}(M))$. 
Motivated by the inferred broadness of the required enhanced spectrum, we parametrize it as a top-hat
\begin{equation}\label{def:PR}
\mathcal{P_R}=\mathcal{A_R}\Theta(k-k_\mathrm{min})\Theta(k_\mathrm{max}-k)\,,
\end{equation}
where the amplitude $\mathcal{A}_{\mathcal{R}}$ and the cutoffs $k_{\min}$, $k_{\max}$ are free parameters to be determined by the joint fit to microlensing and PTA data. 

During the radiation-dominated era, 
the fractional abundance $f \equiv \rho_{\text{PBH}}/\rho_{\text{tot}}$ grows as $\propto a$, thus earlier-formed (smaller) PBHs experience a longer growth. Incorporating this redshift enhancement, the present-day PBH mass function is~\cite{Niemeyer:1997mt,Byrnes:2018clq}
\begin{align}\label{eq:f(M)}
f(M)=\frac{1}{\Omega_\mathrm{dm}}\!
\int\!
\frac{\mathcal{K}/\gamma}{\sqrt{2\pi \sigma_\delta^2}}
e^{-\frac{(\mu^{1/\gamma}+\delta_\mathrm{c})^2}{2\sigma_\delta^2(M_H)}}\mu^{1+\frac1\gamma}\sqrt{\frac{M_\mathrm{eq}}{M_H}}\mathrm{d}\ln M_H,
\end{align}
where $\mu=M/(\mathcal{K}M_H)$. The total PBH abundance is $f_\mathrm{tot}=\int f(M)\mathrm{d}\ln M$, while the mean mass is conventionally introduced as $\langle M\rangle\equiv f_\mathrm{tot}/\int\mathrm{d}\ln M f(M)/M$ to compare with the observational constrains derived under the assumption of a monochromatic mass function. 

\begin{figure}[htbp]
    \centering    \includegraphics[width=0.49\textwidth]{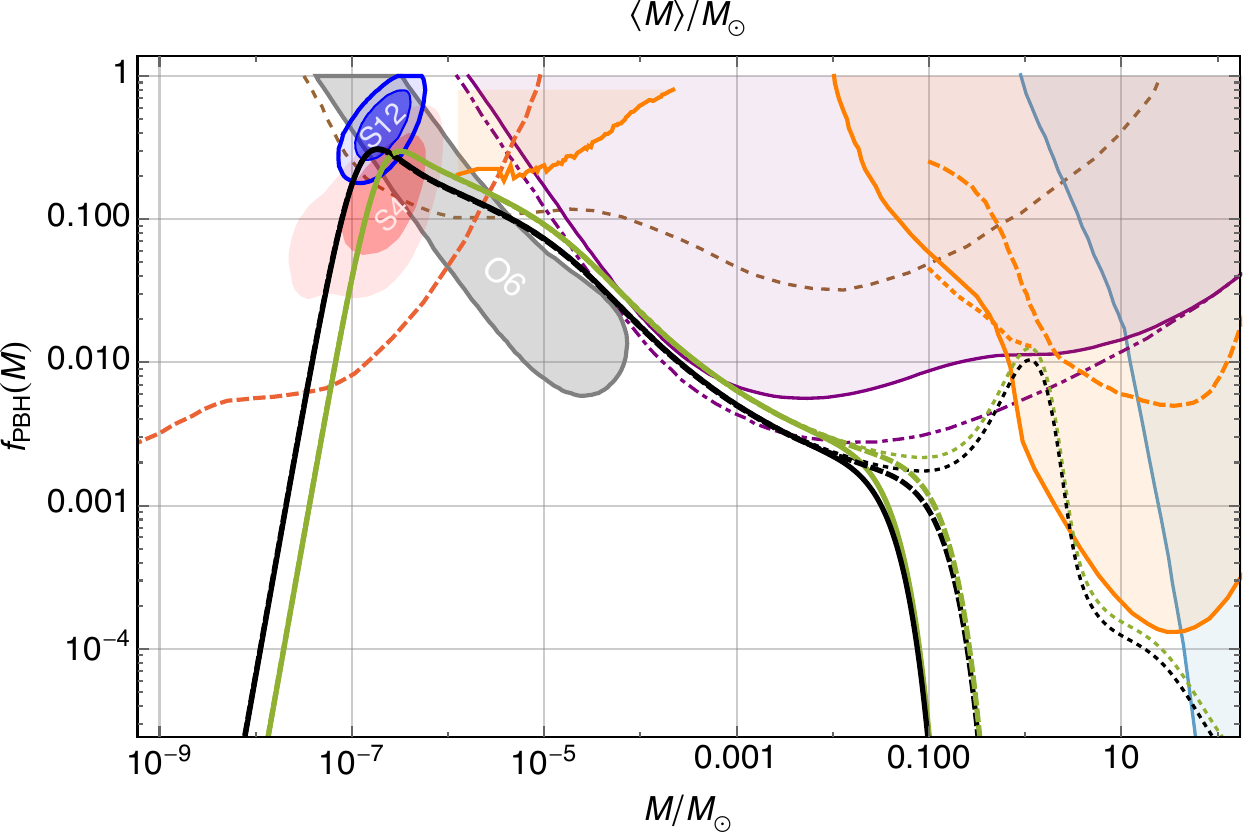}
    
    \vspace{0.5em} 
    \begin{tabular}{c c c c c} 
        \hline 
        $f(M)$ & $\mathcal{A}_{\mathcal{R}}$ 
        & $k_{\mathrm{max}}/\mathrm{Mpc}^{-1}$ 
        & $k_{\mathrm{min}}/\mathrm{Mpc}^{-1}$ 
        & $\langle M\rangle/M_\odot$ \\
        \hline
        \solidblack & $0.03501$ & $3.53\times10^{10}$ & $1.59\times 10^7$ & $3.00\times10^{-7}$ \\
        \dashedblack & $0.03499$ & $3.55\times 10^{10}$ & $8.84\times 10^6$ & $3.00\times10^{-7}$  \\
        \solidsteel & $0.03531$ & $2.69\times 10^{10}$ & $1.59\times 10^7$ & $5.30\times10^{-7}$  \\
        \dashedsteel & $0.03530$ & $2.70\times 10^{10}$ & $8.84\times 10^6$ & $5.30\times10^{-7}$  \\
        \dottedblack & $0.03503$ & $3.51\times 10^{10}$ & --- & $3.00\times10^{-7}$  \\
        \dottedsteel & $0.03533$ & $2.68\times 10^{10}$ & --- & $5.30\times10^{-7}$  \\
        \hline 
    \end{tabular}
\emergencystretch=1.5em
\hyphenpenalty=50
\caption{
The blue and red shaded contours show the 68\% (1$\sigma$) and 95\% (2$\sigma$) C.L. allowed regions of PBH abundance when all 12 microlensing candidates (``S12'') or the 4 secure candidates (``S4'') of Subaru-HSC are attributed to PBHs, respectively \cite{Sugiyama:2026kpv}. The 95\% C.L. allowed region from OGLE 6 events (``O6'') is also shown as the gray shaded region \cite{Niikura:2019kqi}. PBH mass functions are drawn for $f_\mathrm{tot}=1$ with $\langle M\rangle$ set at the 2$\sigma$ blue contour boundaries, while $k_\mathrm{min}$ is fitted to NANO\-Grav data at 2$\sigma$. Mass functions without an IR cutoff are also shown for comparison (thin dotted). Other observational constraints include Subaru-HSC 2019 (light red dashed) \cite{Niikura:2017zjd}, EROS (brown dashed) \cite{EROS-2:2006ryy}, OGLE relaxed (purple shaded) and strict (purple dot-dashed) \cite{Mroz:2024mse}, LIGO asymmetric merger (orange wiggling shaded) \cite{LIGOScientific:2025vwc}, LIGO merger (orange shaded) \cite{Andres-Carcasona:2024wqk}, LIGO stochastic background (orange dashed) \cite{Boybeyi:2024mhp}, LIGO direct search (orange dotted) \cite{Kacanja:2026byy}, and CMB disk accretion (steel blue shaded) \cite{Serpico:2020ehh}.
}\label{fig:pbh}
\end{figure}

The Subaru-HSC results, encompassing all 12 microlensing events, allow for terrestrial-planet-mass PBHs with mean masses in the range $3.00 \times 10^{-7} \, M_\odot \leq \langle M \rangle \leq 5.30 \times 10^{-7} \, M_\odot$ to account for all dark matter ($f_{\mathrm{tot}} = 1$) at the $2\sigma$ confidence level (C.L.). In Fig.~\ref{fig:pbh}, we present two representative mass functions $f(M)$ corresponding to the lower and upper bounds of $\langle M \rangle$ that satisfy $f_{\mathrm{tot}} = 1$ within $2\sigma$ C.L., alongside the OGLE-allowed region and other observational constraints. Consequently, at $2\sigma$ level, any mass function with $f_{\mathrm{tot}} = 1$ must be bounded by these two limiting curves. As $f(M)$ is broad, $f_\mathrm{tot}=1$ implies $f_\mathrm{PBH,max}<1$. A broken power-law approximation given by $f(M) \propto M^{1+1/\gamma}$ for $M < M_{\mathrm{peak}}$ and $f(M) \propto M^{-1/2}$ for $M > M_{\mathrm{peak}}$ gives roughly $f_\mathrm{PBH}(M_\mathrm{peak})\approx(1+\gamma)/(2+3\gamma)\approx0.4$, consistent with numerical integration. Furthermore, the asymmetric tail at high-mass end decays slower than the low-mass side, rendering $\langle M\rangle$ slightly higher than the peak mass $M_\mathrm{peak}$ by a factor of $3/(\gamma+1)\approx2.2$. Nevertheless, both $\langle M\rangle$ and $M_\mathrm{peak}$ are much larger than $M_H(k_\mathrm{max})$ by an $\mathcal{O}(10)$ factor.

As the power spectrum is scale-invariant, $f(M)$ extends toward higher masses as $f(M) \propto M^{-1/2}$, spanning from planet to solar masses while remaining consistent with existing observational bounds. 
When the mass function extends to subsolar-mass regime, the softening of equation-of-state $w$ from QCD transition could substantially enhance $f(M)$, potentially inducing a second peak \cite{Byrnes:2018clq,Carr:2019hud,Garcia-Bellido:2019vlf,Carr:2019kxo,Bodeker:2020stj,DeLuca:2020agl,Magaraggia:2026jhk}. Such a population of subsolar-mass PBHs is of paramount interest for GW astronomy, as it provides a natural bridge to the nanohertz SGWB recently reported by PTA collaborations \cite{NANOGrav:2023hvm,Franciolini:2023pbf,Inomata:2023zup,Wang:2023ost,Liu:2023ymk,Domenech:2024rks,Iovino:2024tyg}.

Both the PBH mass and the frequency of induced GWs are determined by the Hubble scale at horizon reentry of the enhanced curvature perturbation, related through \cite{Saito:2008jc,Sasaki:2018dmp}

\begin{equation}\label{frequency-mass}
    f_{\mathrm{IGW}} \approx 3 \, \mathrm{Hz} \left( \frac{M_{\mathrm{PBH}}}{10^{16} \, \mathrm{g}} \right)^{-1/2}\left( \frac{g_{*s}(\eta_k)}{106.75} \right)^{-1/12},
\end{equation}
where $g_{*s}$ is the number of relativistic degrees of freedom (DOF) for entropy at the reentry moment $\eta_k$, which we set it equal to the DOF in energy density and use the fitting formulas of Ref.~\cite{Saikawa:2018rcs}. Thus, subsolar-mass PBHs are naturally accompanied by substantial production of induced GWs in the nanohertz band. Remarkably, strong evidence for such a SGWB has been reported by PTA experiments \cite{EPTA:2023fyk,EPTA:2023sfo,EPTA:2023xxk,Zic:2023gta,Reardon:2023gzh,Reardon:2023zen,NANOGrav:2023hde,NANOGrav:2023gor,InternationalPulsarTimingArray:2023mzf,Xu:2023wog}, motivating a detailed study of the corresponding induced GWs in the context of planet-mass PBHs.

The resulting induced GW spectrum for Gaussian curvature perturbations $\mathcal{R}$ can be calculated according to \cite{Espinosa:2018eve,Kohri:2018awv,Pi:2020otn,Domenech:2021ztg}
\begin{align}\nonumber
\frac{\Omega_\text{GW}(f)h^2}{1.6\times10^{-5}}
&=\left(\frac{g_{*s}(\eta_k)}{106.75}\right)^{-1/3}\left(\frac{\Omega_{r,0}h^2}{4.1\times10^{-5}}\right)\\\nonumber
&\times 3\int^\infty_0\mathrm{d} v\int^{1+v}_{|1-v|}\mathrm{d} u\frac{\mathscr{T}(u,v)}{4u^2v^2}\mathcal{P}_\mathcal{R}(uk)\mathcal{P}_\mathcal{R}(vk),
\end{align}
where the kernel $\mathscr{T}(u,v)$ is given, e.g., in Ref.~\cite{Pi:2020otn}. $f=k/(2\pi a)$ is the GW frequency. 
Note that such an expression for the induced GW spectrum is computed in the Newton gauge, which has recently been shown to coincide with the strain that a GW detector measures at second order \cite{Domenech:2025ccu}.

Although the induced GW energy spectrum $\Omega_\text{GW}$ can be calculated analytically only for simple spectra—such as monochromatic \cite{Ananda:2006af,Kohri:2018awv,Espinosa:2018eve}, lognormal \cite{Pi:2020otn,Dandoy:2023jot}, or broken power-law \cite{Li:2024lxx}—some general features emerge for sources with finite support. In particular, an infrared $f^3$ scaling \cite{Cai:2019cdl} saturating the causality bound, together with logarithmic corrections \cite{Cai:2018dig,Yuan:2019wwo} (although faded by the dissipation effects \cite{Domenech:2025bvr}), is a generic feature of GW generated during radiation domination \cite{Domenech:2019quo,Domenech:2020kqm,Hook:2020phx,Brzeminski:2022haa,Ghoshal:2023sfa,Cui:2024vws,Domenech:2024rks}. Inside the support $[k_\mathrm{min}, k_\mathrm{max}]$, $\Omega_\mathrm{GW}$ is approximately scale-invariant, mirroring the flat curvature spectrum, while the $f^3$ infrared scaling governs the lower frequencies. This combined profile reproduces the nanohertz signal observed by PTAs.


\textit{Results.}---The parameters $\{\mathcal{A}_{\mathcal{R}}, k_{\max}, k_{\min}\}$ in \eqref{def:PR} can be determined by a joint analysis of microlensing and PTA observations. Given that raw Subaru-HSC data is not public, we infer the posterior of the cutoff scales, $k_{\max}$ and $ k_{\min}$, by combining NANOGrav 15-yr data with existing Subaru-HSC constraints, fixing $\mathcal{A}_{\mathcal{R}}$ to ensure $f_{\mathrm{tot}}=1$. This approach is robust against the marginal statistical uncertainties in $\mathcal{A}_{\mathcal{R}}$ introduced by the microlensing data.


Physically, the parameter space exhibits a clear scale separation. The small uncertainty in $\mathcal{A}_{\mathcal R}$ implies that it is effectively fixed at the level relevant for the NANOGrav sensitivity. The GW signal inferred from NANOGrav is governed primarily by the infrared turnover scale $k_{\min}$, which is decoupled from the small-scale fluctuations responsible for planet-mass PBH formation.

\begin{figure}[htbp]
    \centering
    \begin{subfigure}{0.51\linewidth}
        \centering
        \includegraphics[width=\linewidth]{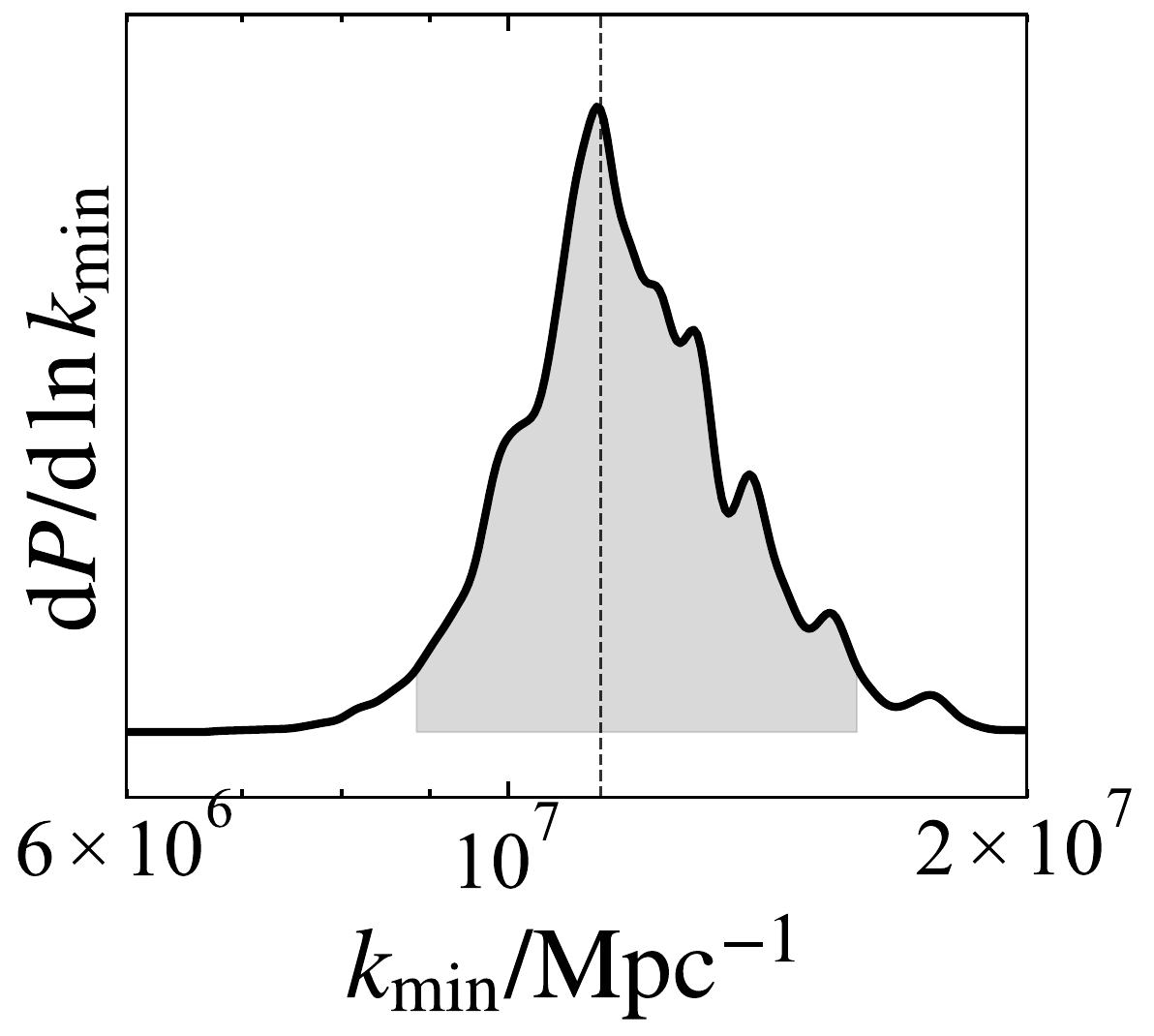}
    \end{subfigure}
    \hfill
    \begin{subfigure}{0.46\linewidth}
        \centering
        \includegraphics[width=\linewidth]{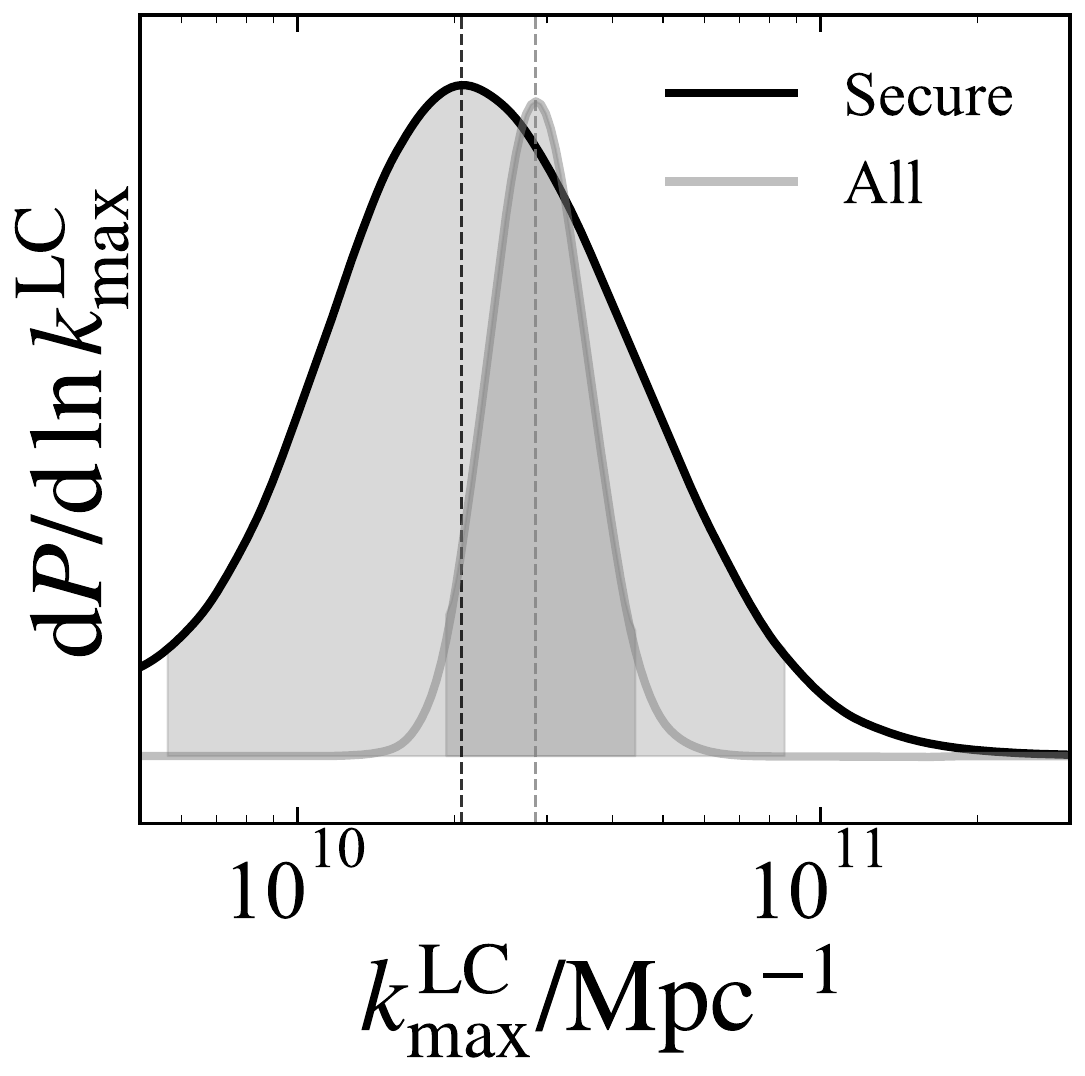}
    \end{subfigure}

    \renewcommand{\arraystretch}{1.35}
    
    \begin{tabular}{|c|c|c|}
    \hline

    \multirow{2}{*}{$k_{\min}/\text{Mpc}^{-1}$} 
    & \multicolumn{2}{c|}{$k^{\rm LC}_{\max}/\text{Mpc}^{-1}$} \\
    \cline{2-3}

    & All & Secure \\
    \hline

    $1.13^{+0.46}_{-0.25}\times10^{7}$ 
    & $2.86^{+1.59}_{-0.95}\times10^{10}$ 
    & $2.06^{+6.44}_{-1.50}\times10^{10}$ \\
    \hline

    \end{tabular}
    \caption{Posterior distributions of $k_{\min}$ (left) and $k^{\rm LC}_{\max}$ (right). The shaded regions indicate the 2-$\sigma$ bounds, and the dashed lines denote the best-fit values. In the right panel, the black curve corresponds to the Subaru-HSC sample with 4 secure candidates, while the gray curve represents the full sample of 12 candidates.}
    \label{fig:posterior}
\end{figure}

Conversely, we derive a conservative estimate for the UV cutoff $k_{\max}^{\rm LC}$, inferred solely from Subaru-HSC microlensing light curves. Note that $k_{\max}^{\rm LC}$ is different from $k_{\max}$ in the preceding section, which also depends on the total abundance. Due to the Poisson likelihood for the observed number of candidates and the imposed condition $f_{\rm tot}=1$, the constraint on the latter is more stringent and biased towards higher values slightly. The resulting posterior distribution of $k_{\min}$ and $k^{\rm LC}_{\max}$ are shown in Fig.~\ref{fig:posterior}. The detailed likelihood construction and prior choices are presented in the \textit{End Matter}. 

The UV cutoff of the GW spectrum is higher than a naive estimate by simply substituting $k_{\max}$ into \eqref{frequency-mass}, because a broad spectrum produces a broad PBH mass function. In particular, the mean mass $\langle M\rangle$ is typically $\mathcal{O}(10)$ times larger than the mass associated with the cutoff scale, which shifts the corresponding GW cutoff frequency upward by a factor of a few. This allows the induced GW signal to reach the sensitivity bands of space-based interferometers such as LISA \cite{LISA:2017pwj}, Taiji \cite{Luo:2019zal}, and TianQin \cite{Luo:2025ewp}.

\begin{figure}[htbp]
    \center
    \includegraphics[width=\linewidth]{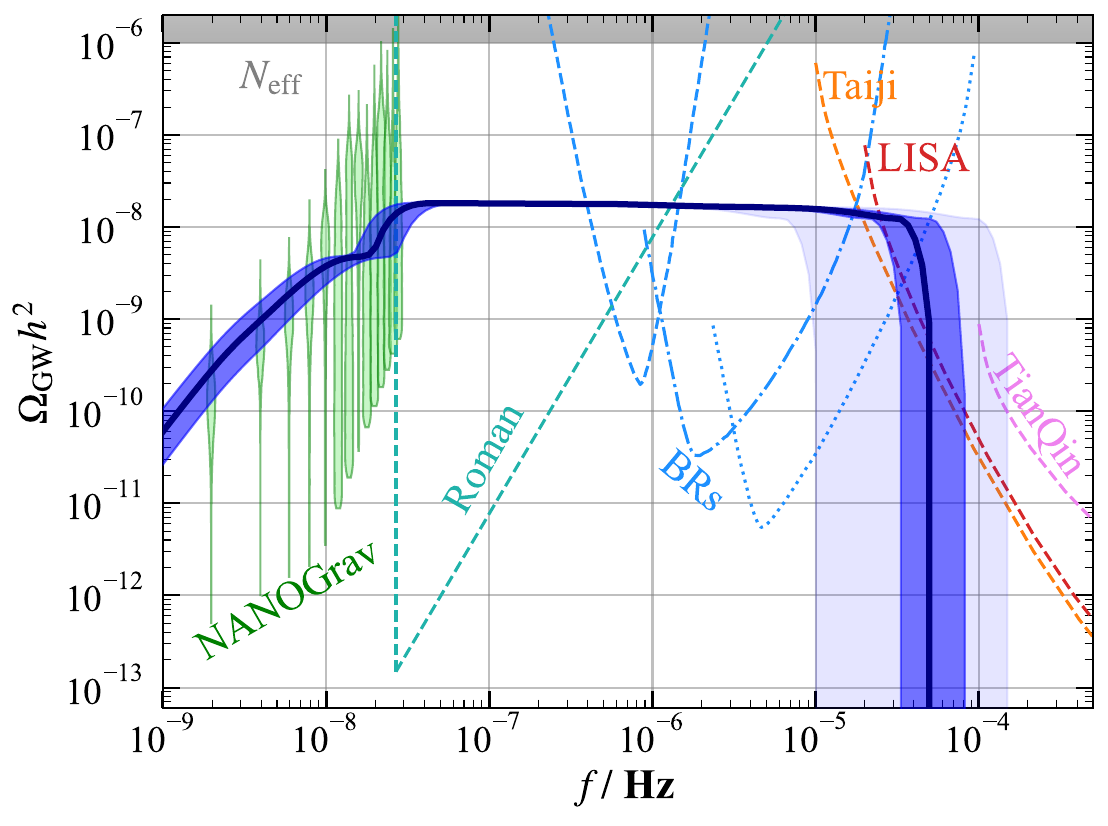}
    \emergencystretch=1.5em
\hyphenpenalty=50
    \caption{GW spectrum $\Omega_\mathrm{GW}$ induced by a top-hat curvature power spectrum. The dark-blue line and shaded region represent the spectrum for the best-fit parameter and the 2-$\sigma$ C.L., respectively, inferred from the NANOGrav 15-yr data \cite{NANOGrav:2023ctt} and the Subaru-HSC sample of all candidates \cite{Sugiyama:2026kpv}, while the light-blue shaded region assumes 4 secure candidates. The first 14 bins of the NANO\-Grav signal \cite{NANOGrav:2023ctt} is shown in green violins together with some future observational projects, including the Roman Telescope (Seagreen) \cite{Wang:2022sxn,Pardo:2023cag}, binary resonances (BRs) (Lightblue) \cite{Foster:2025nzf,Blas:2026xws} shown in LLR (dashed), eLO (dot-dashed), and eSLR (dotted) configurations, LISA (Red) \cite{LISA:2017pwj}, Tai\-ji (Orange) \cite{Luo:2019zal}, and Tian\-Qin (Pink) \cite{Luo:2025ewp}. The gray-shaded region is the bound from relativistic DOF from CMB/BBN \cite{Cyburt:2004yc,Binetruy:2012ze,Planck:2018vyg,Arbey:2021ysg,Grohs:2023voo}.}
    \label{fig:gw}
\end{figure}

In Fig.~\ref{fig:gw}, we present the GW energy spectrum inferred from the posterior parameters, compared with the NANOGrav 15-yr data and the predicted sensitivity curves of some other experiments, including the optimistic Roman \cite{Wang:2022sxn,Pardo:2023cag}, Lunar Laser Ranging (LLR), and Satellite Laser Ranging (SLR) in different orbital configurations, namely the eccentric lunar-orbit (eLO) and highly eccentric Earth-orbit (eSLR) setups \cite{Foster:2025nzf,Blas:2026xws}, as well as LISA \cite{LISA:2017pwj}, Taiji \cite{Luo:2019zal}, and TianQin \cite{Luo:2025ewp} (for 3-yr power-law integrated sensitivity curves with $\mathrm{SNR}=3$~\cite{Thrane:2013oya}).

\textit{Conclusion and Discussions.}---Recent high-cadence observations by Subaru-HSC have revealed a set of ultra-short-timescale microlensing events, for which 
a population of planet-mass primordial black holes provides a viable interpretation where such objects can constitute the totality of dark matter. Given the strong evidence for a nanohertz stochastic gravitational-wave background reported by multiple PTA collaborations, we show that both phenomena can be naturally and simultaneously explained by an enhanced broad top-hat curvature power spectrum extending from $k_{\mathrm{min}}\simeq10^{6}\,\mathrm{Mpc}^{-1}$ to $k_{\mathrm{max}}\simeq3\times10^{10}\,\mathrm{Mpc}^{-1}$ with amplitude $\mathcal{A_R}\simeq3.5\times10^{-2}$. This unified framework not only reconciles the entirety of dark matter, microlensing events, and nanohertz gravitational waves within a single primordial scenario, but also yields a PBH mass function that avoids all existing observational constraints from the planet- to solar-mass range. Notably, the nearly flat induced GW spectrum naturally arising from the proposed top-hat enhancement is associated with a PBH mass function peaking in the planet-mass range, thereby avoiding the PBH overproduction typically encountered during the QCD transition in alternative scenarios. This unified framework is inherently independent of the substantial theoretical uncertainties in the PBH abundance calculation.


This scenario can be tested by multiple upcoming observational projects across different channels. Extended microlensing surveys, including Subaru-HSC, OGLE, and Roman, will significantly improve the statistical precision of short-timescale events, enabling a clear discrimination between PBH and astrophysical lenses. In the GW sector, next-generation PTA projects such as SKA and CPTA will refine the nanohertz spectrum, directly testing the induced GW interpretation. Space-borne interferometers LISA, Taiji, and TianQin cover the milli-Hz band ($10^{-4}$--$10^{-1}$ Hz), and can reach the high-frequency break of the induced spectrum near $\sim10^{-4}$ Hz over multi-year operations. Furthermore, the intermediate frequency gap is bridged by high-precision astrometry from Gaia, Roman ($\sim 10^{-7}$ Hz), and astrophysical resonators such as lunar and satellite laser ranging ($10^{-6}\text{--}10^{-5}$ Hz), providing a complementary search for the predicted flat spectrum. Additionally, the merger of planet-mass PBH binaries presents an intriguing target for high-frequency GW detectors~\cite{Aggarwal:2020olq,Aggarwal:2025noe,Franciolini:2022htd,Goryachev:2021zzn,Campbell:2025mks}. In the coming decades, these multi-messenger probes will decisively test the broad top-hat spectrum framework proposed here, either ruling it out or providing strong support and sharply constraining its parameter space. 

Even if PBHs account for only a few percent of dark matter, as suggested by the subset of four secure Subaru-HSC events, these synergistic observations retain the sensitivity to sharply constrain the total abundance $f_{\mathrm{tot}}$, the mean mass $\langle M\rangle$, and the mass distribution width. In such a regime, the shape-dependence of the observational constraints must be precisely calibrated. Furthermore, while we adopt a benchmark top-hat spectrum to distill the essential phenomenology of the PBH--induced-GW connection, realistic inflationary models typically feature finite power transitions. For instance, the steepest growth of $\mathcal{P_R}$ is generally $k^4$ in single-field inflation \cite{Byrnes:2018txb,Carrilho:2019oqg,Ozsoy:2019lyy,Tasinato:2020vdk,Cole:2022xqc,Artigas:2024ajh}, while the high-$k$ decay remains model-dependent \cite{Atal:2019cdz,Ozsoy:2023ryl}. Such spectral features, particularly those with milder growth, could be further scrutinized by future CMB spectral distortion surveys \cite{Unal:2020mts,Cyr:2023pgw,Tagliazucchi:2023dai,Yi:2023tdk}, providing a vital complementary probe of the primordial power enhancement. We defer a comprehensive analysis of such effects to future work.

We adopted the Press--Schechter formalism to compute the PBH mass function, taking a threshold $\delta_\mathrm{c}\simeq0.48$ derived from an analytical formula given in \cite{Musco:2018rwt,Musco:2020jjb} for a top-hat power spectrum. This threshold is slightly higher than the Harada--Yoo--Kohri limit $\delta_\mathrm{c}\approx0.42$ \cite{Harada:2013epa} widely used in the literature. Conversely, peak-theory analysis indicates that Press--Schechter may systematically underestimate the PBH abundance, suggesting a lower threshold $\delta_\mathrm{cr}\simeq0.31$ \cite{Pi:2024ert}. Although the inferred PBH abundance depends sensitively on the collapse prescription, once the abundance is fixed, the resulting prediction for the induced GW spectrum remains largely insensitive to these different choices. Crucially, with $f_{\mathrm{tot}}$ anchored by Subaru-HSC, the $\mathcal{A}_{\mathcal{R}}$--$k_{\min}$ correlation ensures that any $\delta_c$-induced shift in $\mathcal{A}_{\mathcal{R}}$ is absorbed by $k_{\min}$, maintaining a robust fit to PTA data. This is explicitly shown in the \textit{End Matter}. A similar insensitivity holds \cite{Cai:2018dig,Bartolo:2018rku,Unal:2018yaa} in the presence of non-Gaussianities \cite{Young:2013oia,Harada:2015yda,Biagetti:2018pjj,Passaglia:2018ixg,DeLuca:2019qsy,Biagetti:2021eep,Pi:2021dft,Ferrante:2022mui,Pi:2022ysn,Ballesteros:2024pbe,Caravano:2025diq}. 
We therefore conclude that our main results are not qualitatively affected by these theoretical and statistical uncertainties.
The joint PBH--induced-GW scenario thus emerges as a robust and falsifiable framework, poised to be decisively tested by forthcoming multi-messenger observations.

\textit{Acknowledgements.}---
We thank Misao Sasaki for valuable comments. This work is supported by the National Key Research and Development Program of China Grant No. 2021YFC2203004, and by the DFG under the Emmy-Noether program project number 496592360. 
G.D. and S.P. also acknowledge support by JSPS KAKENHI grant No. JP24K00624. S.P. and A.W. are supported in part by the National Natural Science Foundation of China (NSFC) Grants Nos. 12475066 and 12447101. A.W. is also supported by the UCAS Joint PhD Training Program.


\input{main.bbl}

\clearpage
\setcounter{equation}{0}
\setcounter{figure}{0}
\setcounter{table}{0}
\setcounter{section}{0}
\renewcommand{\theequation}{E\arabic{equation}}
\renewcommand{\thefigure}{E\arabic{figure}}
\renewcommand{\thetable}{E\arabic{table}}
\input{endmatter}

\end{document}

%% file: main.bbl
%

%% file: endmatter.tex

\clearpage
\onecolumngrid

\pagestyle{plain}

\begin{center}
    {\Large\bfseries End Matter}
\end{center}

\twocolumngrid

\section{Equation of state}
In the early universe, the equation-of-state parameter $w \equiv p/\rho$ 
characterizes the ratio of pressure to energy density. During most of the radiation-dominated era, the cosmic plasma behaves approximately as an ideal relativistic fluid with $w \simeq 1/3$. However, as the temperature decreases and particle species become nonrelativistic, the pressure becomes partially suppressed relative to the energy density and $w$ temporarily deviates from $1/3$. This effect is particularly pronounced at temperatures around $200\,\mathrm{MeV}$, when the universe undergoes the QCD transition from a quark--gluon plasma to a hadronic phase. The evolution of $w$ throughout the thermal history of the universe has been computed using lattice QCD and Standard Model thermodynamics in, \textit{e.g.} \cite{Byrnes:2018clq,Saikawa:2018rcs,Carr:2019kxo}. In this \textit{Letter} we will use the $w(T)$ given in Ref. \cite{Carr:2019kxo}. To find out the change of $w$ as a function of PBH horizon mass, we neglect the corrections to the temperature, and assume $T\propto1/a$ thoughout the radiation-dominated era, which gives $M_H/M_\mathrm{eq}=\big(T/T_\mathrm{eq}\big)^{-1/2}$, 
where $M_\mathrm{eq}\approx2.9\times10^{17}M_\odot$ and $T_\mathrm{eq}\approx8.0\times10^{-4}~\mathrm{MeV}$ are the horizon mass and temperature at matter-radiation equality. Then we can write $w(T)$ as a function of horizon mass $w(T(M))$. 

\begin{figure}[htbp]
    \center
    \includegraphics[width=\linewidth]{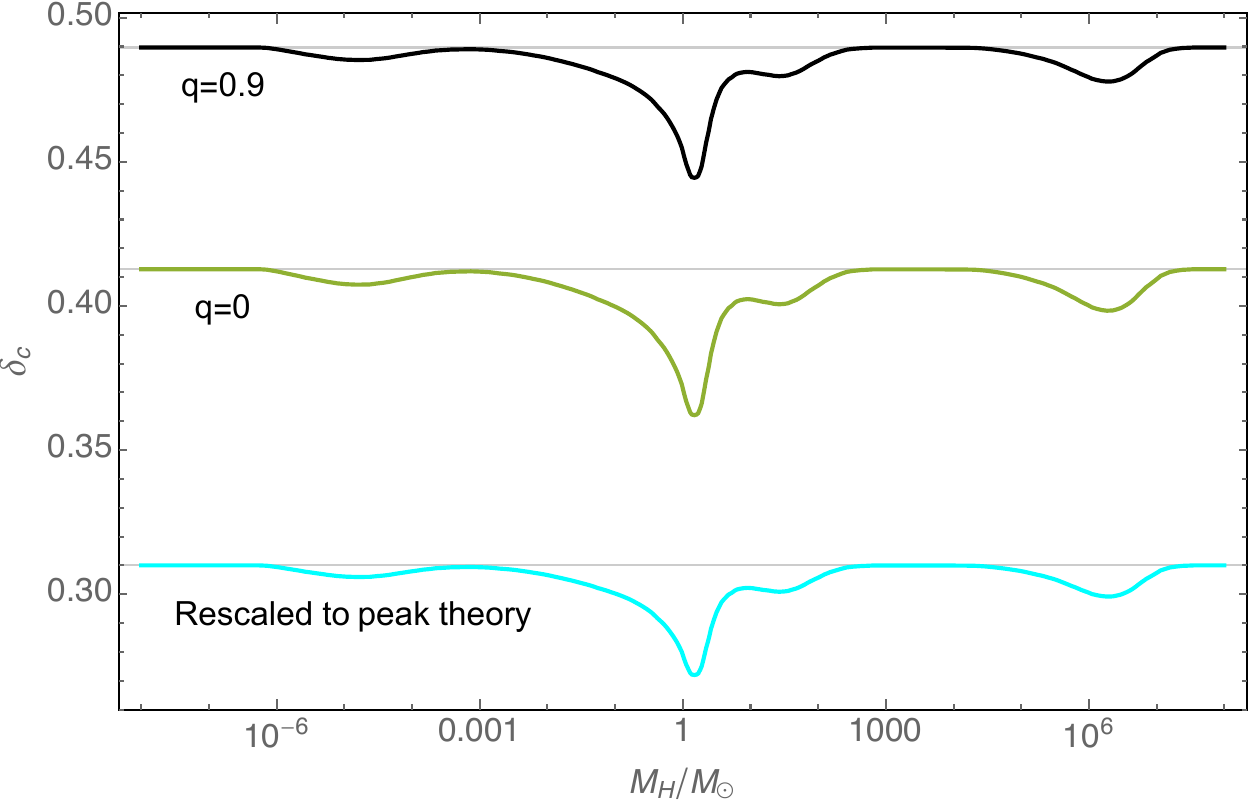}
    \caption{The threshold of density contrast $\delta_\mathrm{c}(M)$ for different shape factor $q$, as a function of horizon mass $M_H$. See the text for details. }
    \label{fig:deltac}
\end{figure}

More PBHs can be produced when the threshold $\delta_\mathrm{c}(w)$ decreases as the equation-of-state parameter $w$ is reduced. The threshold $\delta_\mathrm{c}(w,q)$ depends on both the equation-of-state parameter and the shape parameter $q\equiv -r_m^2\mathcal{C}''(r_m)/4\mathcal{C}(r_m)$, which roughly describes the width of the compaction function $\mathcal{C}\equiv 2G\delta M/(re^{\mathcal{R}(r)})$ at its maximum $r_m$. (See \cite{Escriva:2019phb} for details.) Ref. \cite{Escriva:2020tak} derived an analytical relation of $\delta_\mathrm{c}(w,q)$ based on a fiducial set of curvature profiles, of which the shape factor can be calculated once the spectral shape is given. 
In this \textit{Letter}, we follow the argument in Ref. \cite{Musco:2020jjb}, which gives $q=0.9$ and $\delta_\mathrm{c}\approx0.48$ for a top-hat power spectrum. We note that the analytical formula in \cite{Escriva:2019phb} is consistent with Ref. \cite{Musco:2020jjb}, as $\delta_\mathrm{c}(w=1/3,q=0.9)\approx0.4897$. Therefore, we use the analytic formula $\delta_\mathrm{c}(w,q)$ throughout our calculation.

Finally, by substituting $w(T(M))$ into $\delta_\mathrm{c}(w,q=0.9)$, we get $\delta_\mathrm{c}(w(T(M)))$, which is the black curve in Fig. \ref{fig:deltac}. This is what we used when calculating the PBH mass function \eqref{eq:f(M)}. We ignore the slight difference between $M_H$ and $M_\mathrm{PBH}$ in the argument of $\delta_\mathrm{c}(w(T(M))$.

\section{Robustness against the PBH threshold}
In the main text, we mention that our conclusion is insensitive to the method we choose to calculate the PBH abundance. In this Section we show it explicitly by choosing different thresholds. As is discussed, we use the analytical formula $\delta_\mathrm{c}(w,q)$ proposed in \cite{Escriva:2020tak}, with $q=0.9$ for a top-hat spectrum, justified in \cite{Musco:2020jjb}. In some literature, the shape dependence in the threshold is ignored, and $\delta_\mathrm{c}$ goes to the Harada-Yoo-Kohri limit $\delta_\mathrm{c}=0.41$ \cite{Harada:2013epa}, which is consistent which $\delta_\mathrm{c}(w=1/3,q=0)=0.412$. In peak theory (PT), although the top-hat spectrum has not been studied, an analysis on broad log-normal spectra suggests that $\delta_\mathrm{c}=0.31$ for radiation dominated era \cite{Pi:2024ert}. Unfortunately, there is no general formula for $\delta_\mathrm{c}(w,q)$ in peak theory. So we simply rescale $\delta_\mathrm{c}$ as
\begin{equation}\label{def:deltaPT}
    \delta_\mathrm{c}^\mathrm{(PT)}(w)\approx\frac{0.31}{\delta_\mathrm{c}(1/3,0)}\delta_\mathrm{c}(w,0)
\end{equation}
to mimic the threshold in peak theory for an arbitrary $w$.

\begin{figure}[htbp]
    \centering
    \includegraphics[width=0.9\linewidth]{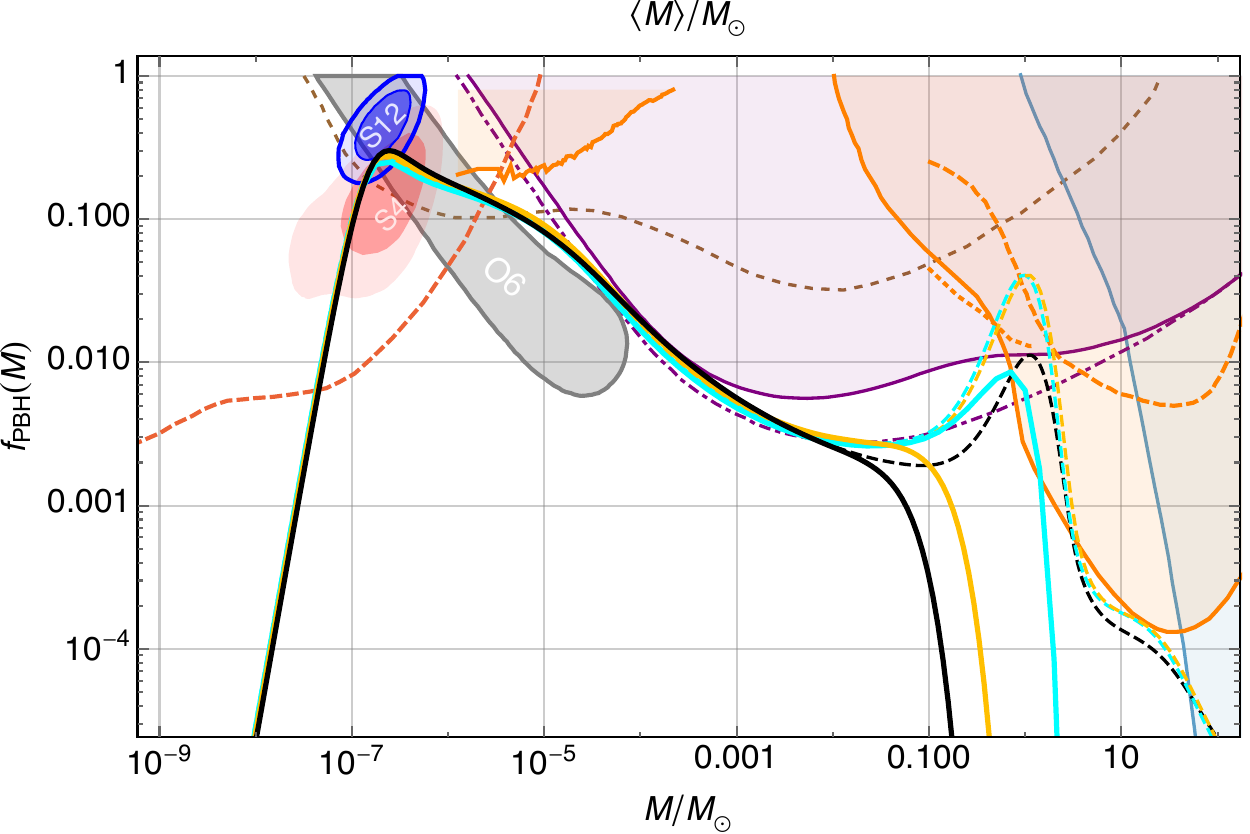}
    \vspace{1ex}
    \includegraphics[width=0.9\linewidth]{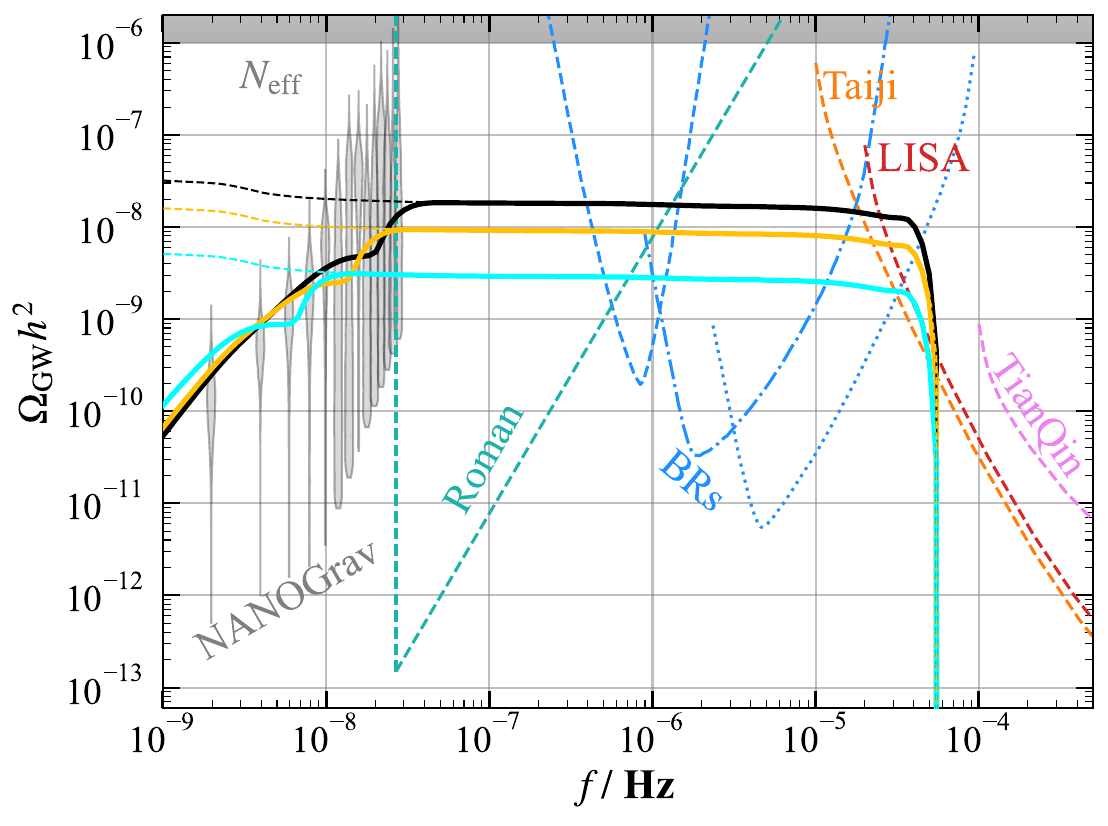}
    \begin{tabular}{c c c c c}
        \hline
        curve & $\delta_{\mathrm{c}}^{(w=1/3)}$ & $\mathcal{A}_{\mathcal{R}}$ & $k_\mathrm{max}/\mathrm{Mpc}^{-1}$ & $k_{\mathrm{min}}/\mathrm{Mpc}^{-1}$ \\
        \hline
        \solidblack & 0.490 & $0.0352$ & $3.08\times10^{10}$ & $1.19\times10^{7}$ \\
        \solidorange & 0.412 & $0.0250$ & $3.08\times10^{10}$ & $8.15\times 10^{6}$ \\
        \solidcyan & 0.310 & $0.0141$ & $3.00\times10^{10}$ & $3.80\times 10^{6}$ \\
        \dashedblackthin & 0.490 & $0.0352$ & $3.08\times10^{10}$ & ---\\
        \dashedorange & 0.412 & $0.0249$ & $3.14\times 10^{10}$ & --- \\
        \dashedcyan & 0.310 & $0.0141$ & $3.02\times10^{10}$ & --- \\
        \hline
    \end{tabular}
    \caption{PBH mass function (upper panel) and induced GW spectrum (lower panel) with $f_\mathrm{tot}=1$ and $\langle M\rangle=4\times10^{-7}~M_\odot$, for different threshold $\delta_\mathrm{c}$. The parameters are shown in the table. In PTA data, there is a pronounced correlation between the power spectrum amplitude $\mathcal{A}_{\mathcal R}$ and the IR cutoff $k_{\min}$, which are adjusted accordingly for different $\delta_\mathrm{c}$ (see text for details). The thin dotted curves are the mass functions and induced GW spectra without an IR cutoff, which exhibit strong tension with LVK constraints and PTA observations, respectively.}\label{fig:pbh2}
\end{figure}

The resulting mass functions and induced GW power spectra for different $\delta_\mathrm{c}$ are shown in Fig. \ref{fig:pbh2}. We see that for a smaller threshold of $\delta_\mathrm{c}$, the curvature perturbation $\mathcal{A_R}$ is less enhanced to generate the same amount of PBHs. Thus the amplitude of the induced GW is smaller, which implies a smaller IR cutoff frequency due to the correlation between the GW amplitude and $f_\mathrm{min}$ in the PTA data, as is clearly shown in the lower panel of Fig. \ref{fig:pbh2}. Such a smaller IR cutoff in turn renders the PBH mass to extend further, down to the subsolar- to solar-mass range which experiences a secondary mass peak due to the QCD transition. This peak might be observed in the very near future by direct or indirect searches in the LVK data. Nevertheless, the main part of the mass function from $10^{-7}~M_\odot$ to $0.1~M_\odot$, is robust against the thresholds and the methods of calculating, which can be tested by the future experiments we discussed in the \textit{Letter}.

\section{Likelihood construction}

We perform a Bayesian inference on the parameter vector $\theta = (\mathcal{A}_{\mathcal R}, k_{\rm min}, k_{\rm max})$ by combining NANOGrav 15-yr PTA data with Subaru HSC microlensing constraints. Since the observables are statistically independent, the joint likelihood factorizes as:
\begin{equation}
    \mathcal{L}(\mathcal{A}_{\mathcal R},\,k_{\rm min},\,k_{\rm max})=\mathcal{L}_{\rm PTA}(\theta)\times \mathcal{L}_{\rm HSC}(\theta).
\end{equation}
$\mathcal{L}_{\rm PTA}(\theta)$ is evaluated using the first 14 frequency bins of the NANOGrav data via the \texttt{PTArcade} code \cite{andrea_mitridate_2023,Mitridate:2023oar}. For the microlensing component, we approximate $\mathcal{L}_{\rm HSC}(\theta)$ using the monochromatic-PBH likelihood evaluated at the total abundance $f_{\rm tot}$ and average mass $\langle M \rangle$, neglecting the effects of the extended mass function.

For a top-hat curvature power spectrum, the observables depend on different subsets of parameters. For the GW sector, the present-day energy density induced by a top-hat curvature spectrum can be approximated as:
\begin{align}
    \Omega_{\rm{GW},0}&h^2(f)\approx1.6\times10^{-6} \left(\frac{g_{*s}(f)}{106.75}\right)^{-1/3}\mathcal{A}_\mathcal{R}^2\notag\\
    &\times\left\{
    \begin{matrix*}[l]
        \mathcal{I}(f/f_{\rm min}) & f\lesssim f_{\rm min}\\
        0.82  & f_{\rm min}\ll f\ll f_{\rm max}\\
        \mathcal{U}(f/f_{\rm max}) & f\gtrsim f_{\rm max}
    \end{matrix*},
    \right.
\end{align}
where $g_{*s}(f)$ denotes the number of relativistic DOF of entropy density at horizon reentry. The prefactor accounts the dilution of GW compared to the background, while $\mathcal{I}$ and $\mathcal{U}$ characterize the IR and UV tails of the spectrum, respectively. Since PTA observations probe the low-frequency tail of the induced gravitational-wave spectrum, they are primarily sensitive to $\mathcal{A}_{\mathcal R}$ and $k_{\rm min}$. In contrast, as shown in Fig.~\ref{fig:pbh}, the PBH abundance and characteristic mass are determined by fluctuations at the smallest scales, thus depending on $\mathcal{A}_{\mathcal R}$ and $k_{\rm max}$. 

Owing to the exponential sensitivity of the PBH abundance to the perturbation amplitude, the Subaru HSC constraints effectively fix $\mathcal{A}_{\mathcal R}$, making its marginalization in the PTA likelihood unnecessary. Meanwhile, we note that the ratio between the mean PBH mass and the horizon mass at the cutoff scale, $\alpha \equiv \langle M\rangle / M_H(k_{\max})$, remains nearly constant within the parameter space of interest, typically $\alpha \simeq 30\text{--}31$. We therefore directly infer the posterior of $k_{\max}$ from the microlensing light-curve likelihood,
\begin{equation}
    \mathcal{L}_{\rm HSC}(\theta)\approx \mathcal{L}^{\rm LC}_{\rm HSC}(\alpha M_H(k_{\max})),
\end{equation}
thereby avoiding the substantial theoretical uncertainties associated with PBH formation prescriptions. In our analysis, we adopt a fixed value $\alpha = 30$ and denote the corresponding ultraviolet cutoff as $k_{\max}^{\rm LC}$.

Then, since the precise value of $\mathcal{A}_{\mathcal R}$ is not critical and is already sufficiently constrained, we fix it to a benchmark value $\mathcal{A}_{\mathcal R}=3.5\times 10^{-2}$ and adopt uniform priors on $\ln k_{\min}$ and $\ln k_{\max}^{\rm LC}$. As $k_{\min}$ and $k^{\rm LC}_{\max}$ affect different observables and are effectively decoupled, their posterior distributions are inferred independently. The resulting marginalized posteriors are shown in Fig.~\ref{fig:posterior}.